\documentclass[aps,prb,twocolumn,showpacs,letterpaper,superscriptaddress]{revtex4-1}
\usepackage{amsmath}
\usepackage{comment}
\usepackage{color}
\usepackage[usenames,dvipsnames]{xcolor}
\usepackage{soul} 
\usepackage{graphics}\usepackage{epsfig}\usepackage{subfigure}
\usepackage{hyperref} 
\usepackage{multirow}
\usepackage{amsmath}

\newcommand{\Tc}{T$_{\rm C}$}

\begin{document}

\title{Why Mercury is a Superconductor}
\author{Cesare Tresca}
\affiliation{Department of Physical and Chemical Sciences,  University of L'Aquila, Via Vetoio 10, I-67100 L'Aquila, Italy} 
\affiliation{Dipartimento di Fisica, Sapienza Universit\`a  di Roma, 00185 Roma, Italy} 
\author{Gianni Profeta}
\affiliation{Department of Physical and Chemical Sciences,  University of L'Aquila, Via Vetoio 10, I-67100 L'Aquila, Italy} 
\affiliation{SPIN-CNR, University of L'Aquila, Via Vetoio 10, I-67100 L'Aquila, Italy}
\author{Giovanni Marini}
\affiliation{Department of Physical and Chemical Sciences,  University of L'Aquila, Via Vetoio 10, I-67100 L'Aquila, Italy} 
\author{Giovanni B. Bachelet}
\affiliation{Dipartimento di Fisica, Sapienza Universit\`a  di Roma, 00185 Roma, Italy}
\author{Antonio Sanna}
\affiliation{Max-Planck-Institut f\"ur Mikrostrukturphysik, Weinberg 2, D-06120 Halle, Germany}
\author{Matteo Calandra}
\affiliation{Sorbonne Universit\'e, CNRS, Institut des Nanosciences de Paris, UMR7588, F-75252, Paris, France}
\affiliation{Department of Physics, University of Trento, Via Sommarive 14, 38123 Povo, Trento, Italy}
\author{Lilia Boeri}
\affiliation{Dipartimento di Fisica, Sapienza Universit\`a di Roma, 00185 Roma, Italy} 
\begin{abstract}
{ Despite being the oldest known superconductor, solid mercury is mysteriously
absent from all current computational databases of superconductors. 
In this work, we present a critical study of its superconducting properties based on state-of-the-art superconducting density-functional theory. Our calculations reveal numerous anomalies in electronic and lattice properties,  which can
mostly be handled, with due care, by modern ab-initio techniques.
In particular, we highlight an anomalous role of (i) electron-electron correlations on structural properties (ii) spin-orbit coupling on the dynamical stability, and (iii) semicore $d$ levels on the effective Coulomb interaction and, ultimately, the critical temperature.
}

\end{abstract}
\pacs{75.70.Tj, 
 74.20.Pq, 
 74.25.Dw 
}
\maketitle

\section{Introduction}
In 1911 Kamerlingh Onnes,\cite{KOnnes_Hg} investigating the transport properties of mercury at low temperatures, observed for the first time a superconducting (SC) transition: below a critical temperature \Tc = 4.15~K, the electrical resistivity dropped to zero. The discovery marked a milestone in physics history. 
The first microscopic theory of 
this phenomenon was formulated only fifty years later by Bardeen, Cooper, and Schrieffer (BCS).\cite{PhysRev.108.1175} Their theory, refined  through the Migdal--\'Eliashberg (ME)\cite{Migdal_1958,Eliashberg_1960} Green's function formalism and the Morel and Anderson Coulomb pseudopotential $\mu^*$,\cite{MorelAnderson} permitted to draw an accurate picture of
the normal and the SC phase of conventional (phonon-mediated) superconductors.

In the 60's and 70's, when an {\it ab-initio} solution of the  
\'Eliashberg equations was beyond available computational capabilities,
mercury, among others, served as a benchmark to derive approximate analytical expressions for various superconducting properties, 
whose main  ingredients were extracted from  experiments. 
Normal-state electronic structure was inferred from de Haas van Alphen,\cite{PhysRev.148.644} magnetoresistance, and cyclotron-resonance measurements;\cite{PhysRev.166.728,PhysRev.175.928} phonon dispersion curves from neutron inelastic  scattering;\cite{doi:10.1080/00150197708237132} while the  \'Eliashberg function  $\alpha^2F(\omega)$ and the SC gap
 from tunneling experiments.\cite{PhysRev.135.A306,PhysRev.188.716} 
Notable examples are the McMillan-Allen-Dynes\cite{PhysRev.167.331,PhysRevB.12.905} approximate formulas for \Tc.

Towards the end of the century,  progress in Density Functional (Perturbation) Theory~\cite{Baroni2001}
allowed first-principles calculations of the electron-phonon spectral function\cite{Savrasov_1996}, 
superconducting \Tc's and gaps.\cite{GiustinoRevModPhys2017, SPG_SCDFTfunctional_PRL2020} 
These methods, combined
with modern crystal-structure prediction algorithms,~\cite{Pickard2011,Curtarolo2013,PhysRevB.99.054102,Oganov2019, FloresLivas2020} and with Ashcroft's intuition of high-\Tc\  SC in hydrogen-rich metallic alloys,\cite{Ashcroft_2004} were the driving force behind the {\em hydride rush} of the last five years.~\cite{FloresLivas2020,Roadmap}

Following these achievements, Density Functional Theory (DFT) based methods 
are rapidly becoming the tool of choice to guide new superconductor discoveries.
The field is evolving in the direction of high-throughput material design; this requires extensive benchmarks on known materials
to vouch the accuracy of current theoretical/computational methods.
The first results seem encouraging: for most conventional superconductors, the agreement between different theoretical approaches and experiment is remarkable.\cite{Savrasov_1996,PhysRevB.101.134511, SPG_SCDFTfunctional_PRL2020}
However, a few notable exceptions exist.
In particular, a close inspection of the available literature reveals that mercury is unexplicably absent from all currently accessible benchmark calculations.

This paper reports our attempt to fill this gap. In particular,
we address the following questions: if Onnes had not discovered superconductivity in mercury, could we predict it today? And, even more importantly, can state-of-the-art theoretical and computational approaches completely describe superconductivity in mercury?
We will show that the answers are not straightforward, 
since, in mercury, all physical properties relevant for conventional
superconductivity,
i.e. the electronic structure, phonon dispersions, electron-phonon coupling and Coulomb matrix elements, are anomalous in some respect.

 In the following, we will discuss each of these aspects separately, 
 and show how 
 they concur to determine a consistent picture of superconductivity in this fascinating element.

\section{Results and discussion}

\begin{figure}[b!]
\centering
\includegraphics[width=0.475\linewidth]{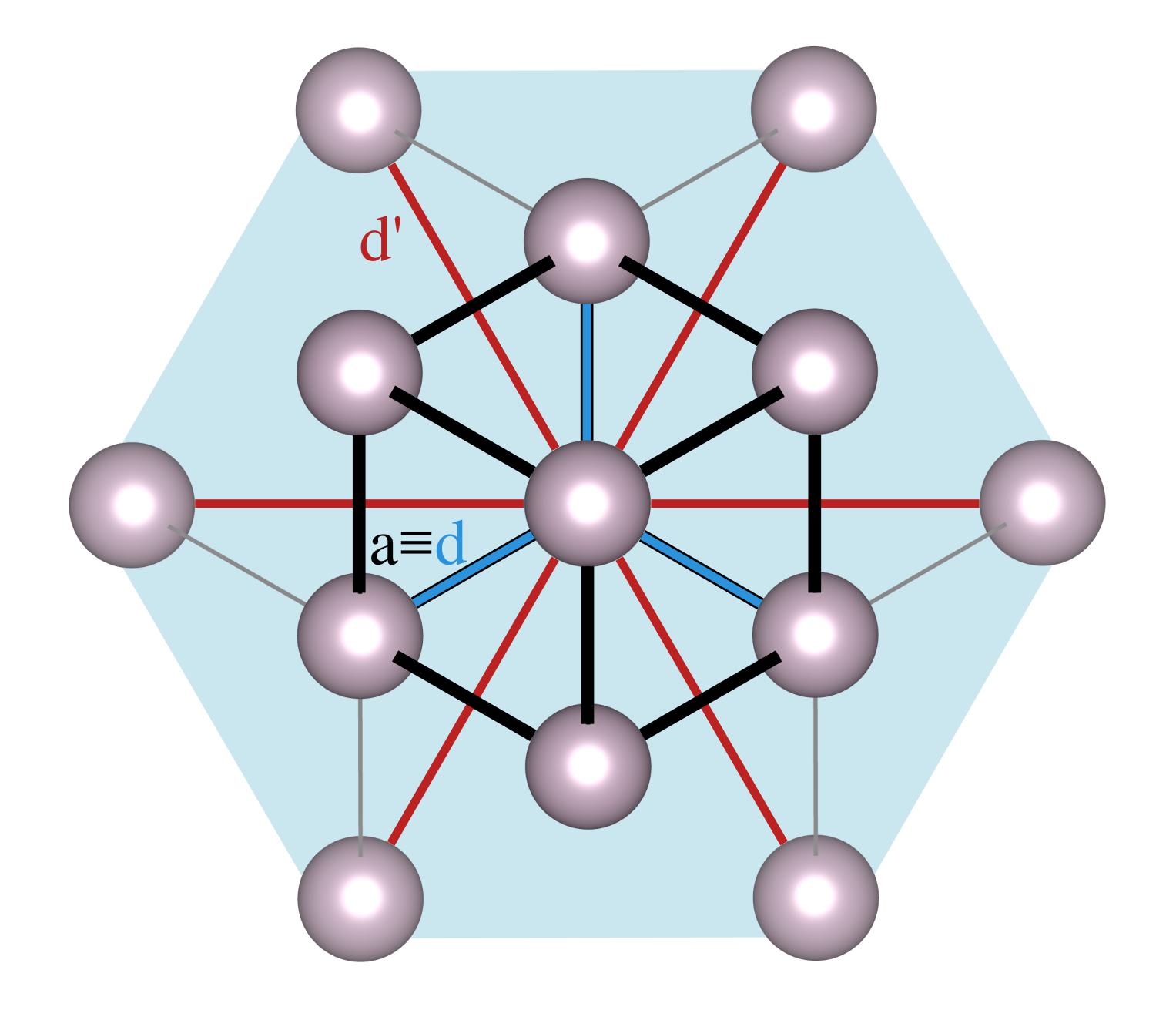}~\includegraphics[width=0.475\linewidth]{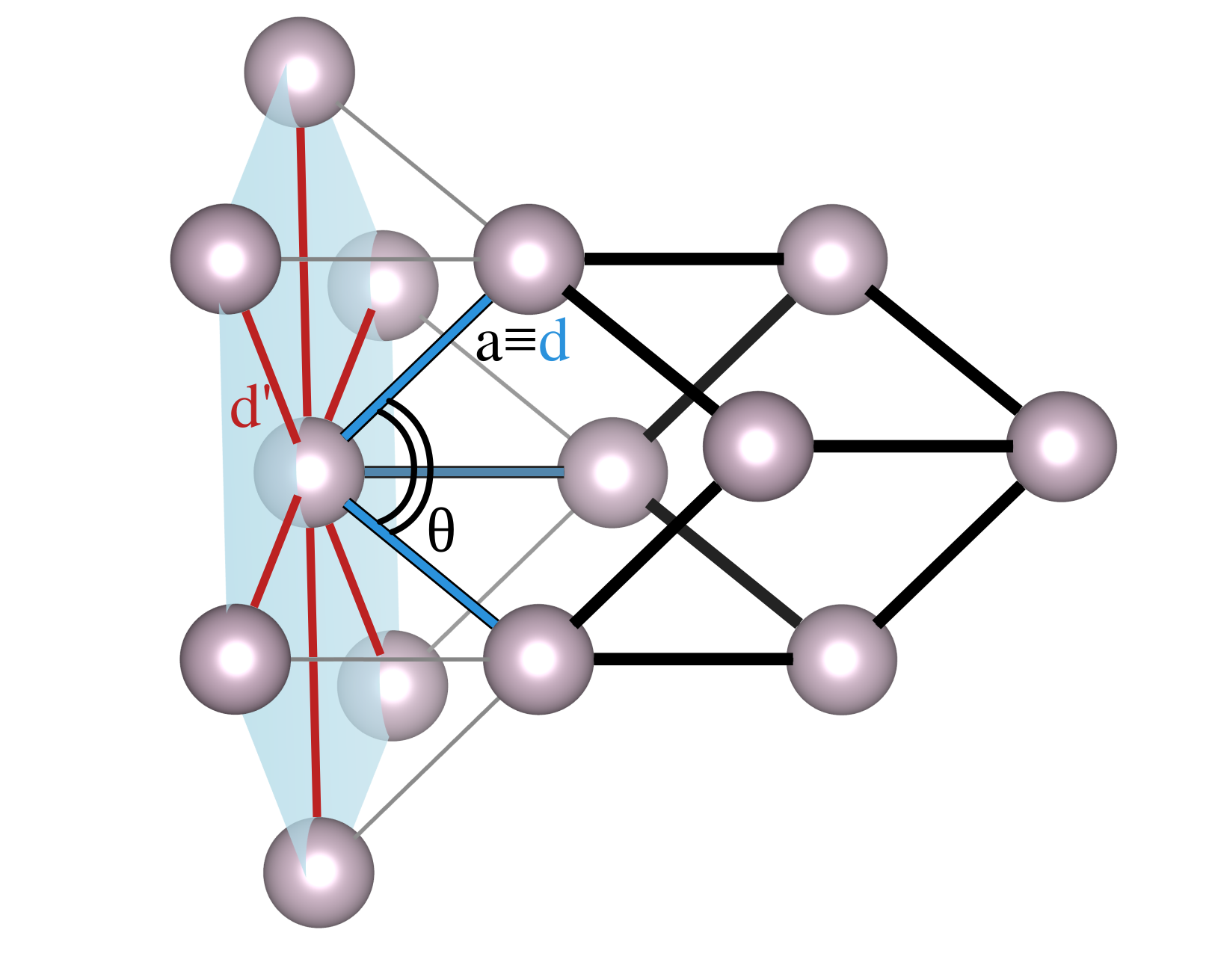}
\caption{Different views of the experimental $\alpha$-Hg structure of superconducting solid mercury in the monoatomic rhombohedral lattice (space group $R3m$ or 160~\cite{Pauling1947,doi:10.1107/S0365110X57000134,CRC1997}). The three primitive lattice vectors are indicated by blue lines;  the angle between each pair of them, $\theta$ , equals to 60$^{o}$ in the fcc structure, but deviates from this value in the rhombohedral phase.}
\label{fig1}
\end{figure}

{\bf Crystal Structure: }
At ambient conditions mercury is liquid, but below 235~K it crystallizes in a monoatomic rhombohedral lattice, the so-called $\alpha$ phase, ~\cite{Pauling1947,doi:10.1107/S0365110X57000134,CRC1997}
shown in Fig. \ref{fig1}, which
is commonly accepted as the actual superconducting phase of mercury. ~\cite{PhysRevB.48.14009,PhysRev.111.82,Pauling1947,doi:10.1107/S0365110X57000134,BRANDT196518,PhysRev.152.548,doi:10.1080/00150197708237132, CRC1997,PhysRevB.48.14009, PhysRevB.53.569,PhysRevB.48.14009, PhysRevB.53.569,MORIARTY198841,Biering2011}

 The structure may be seen as the compression of the fcc structure along a three-fold axis, causing the six equatorial distances ($d^\prime$, red in Fig.\ref{fig1}) to become greater than the six others ($d$, blue in Fig.\ref{fig1}).\cite{Pauling1947} The interatomic distances $d$ and $d'$
 are uniquely determined by the lattice parameter, $a$ and the rhombohedral  angle, $\theta$, which would equal 60$^{\rm o}$ in the undistorted fcc structure.
Total energy structural optimization using  
Generalized Gradient Approximation (GGA)\cite{PhysRevLett.77.3865} functional yields $a_{ }=d_{ }=3.12$~\AA~and $\theta_{ }=78.51^o$ ($d^\prime_{ }=4.83$~\AA)\footnote{The Local Density Approximation (LDA) gives in $a_{LDA}=d_{LDA}=2.94$~\AA~and $\theta_{LDA}=74.54^o$ ($d^\prime_{LDA}=4.68$~\AA)}, 
to be compared with the experimental values $a$=$d$=3.005~\AA~and $\theta=70.53^o$\cite{CRC1997} ($d^\prime =4.907$ ~\AA).

The $\sim 5 \%$ discrepancy between theory and experiment, already reported in literature, signals that semi-local energy functionals cannot properly reproduce the experimental lattice  parameters.\cite{PhysRevB.74.094102, melting}
A  $\sim 5 \%$ accuracy on the structural parameters would be considered acceptable for most materials.
However, in  mercury even minor structural differences cause dramatic effects on the electronic and dynamical properties, which are instead perfectly
reproduced assuming the experimental lattice structure. Hence, in the following, all calculations of the electronic, dynamical and superconducting properties will employ the experimental lattice crystal structure, and the PBE-GGA
exchange and correlation functional.
Further computational details can be found in the Appendix.

\begin{figure}[]
\centering
\includegraphics[width=0.95\linewidth]{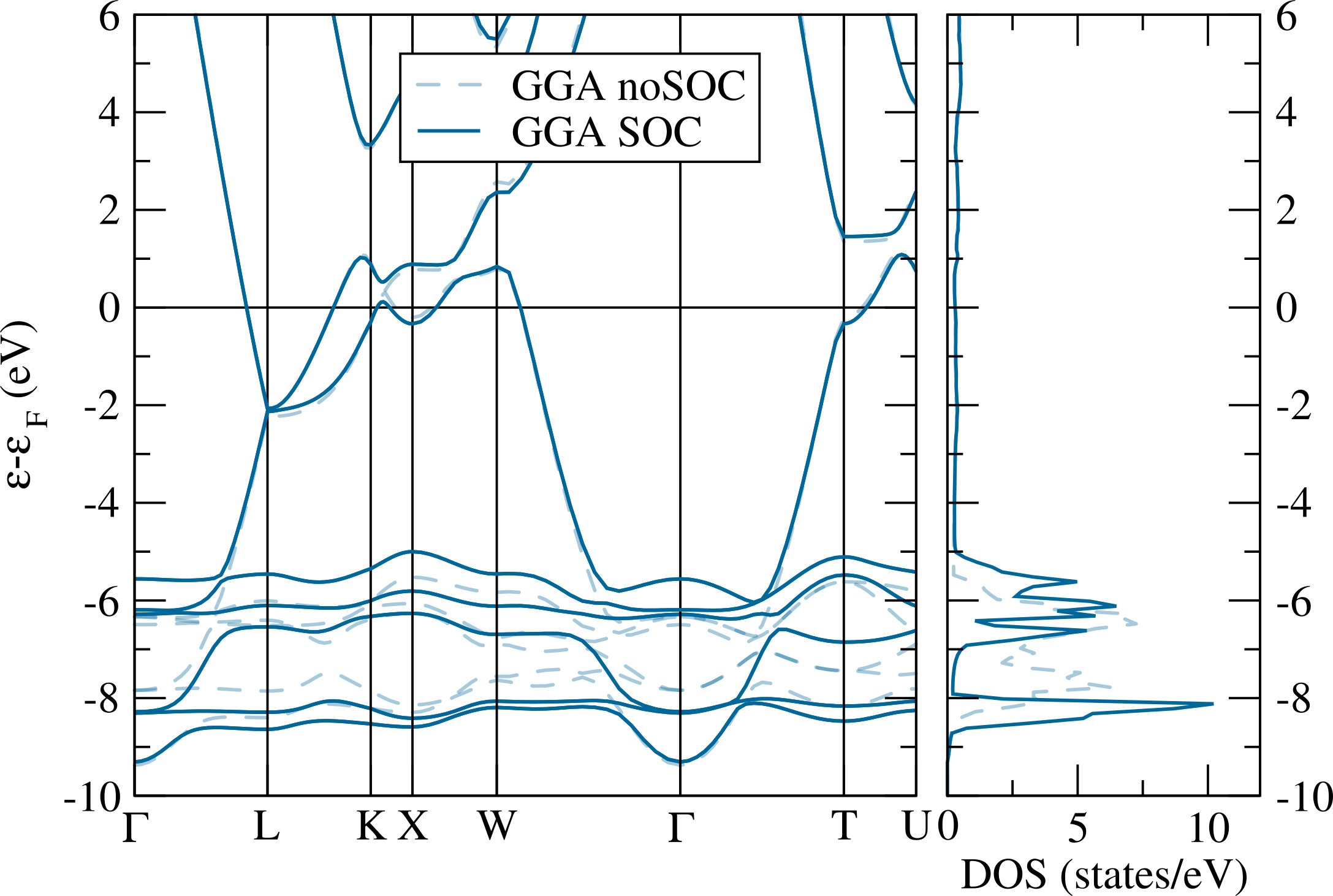}\\\qquad\includegraphics[width=0.425\linewidth]{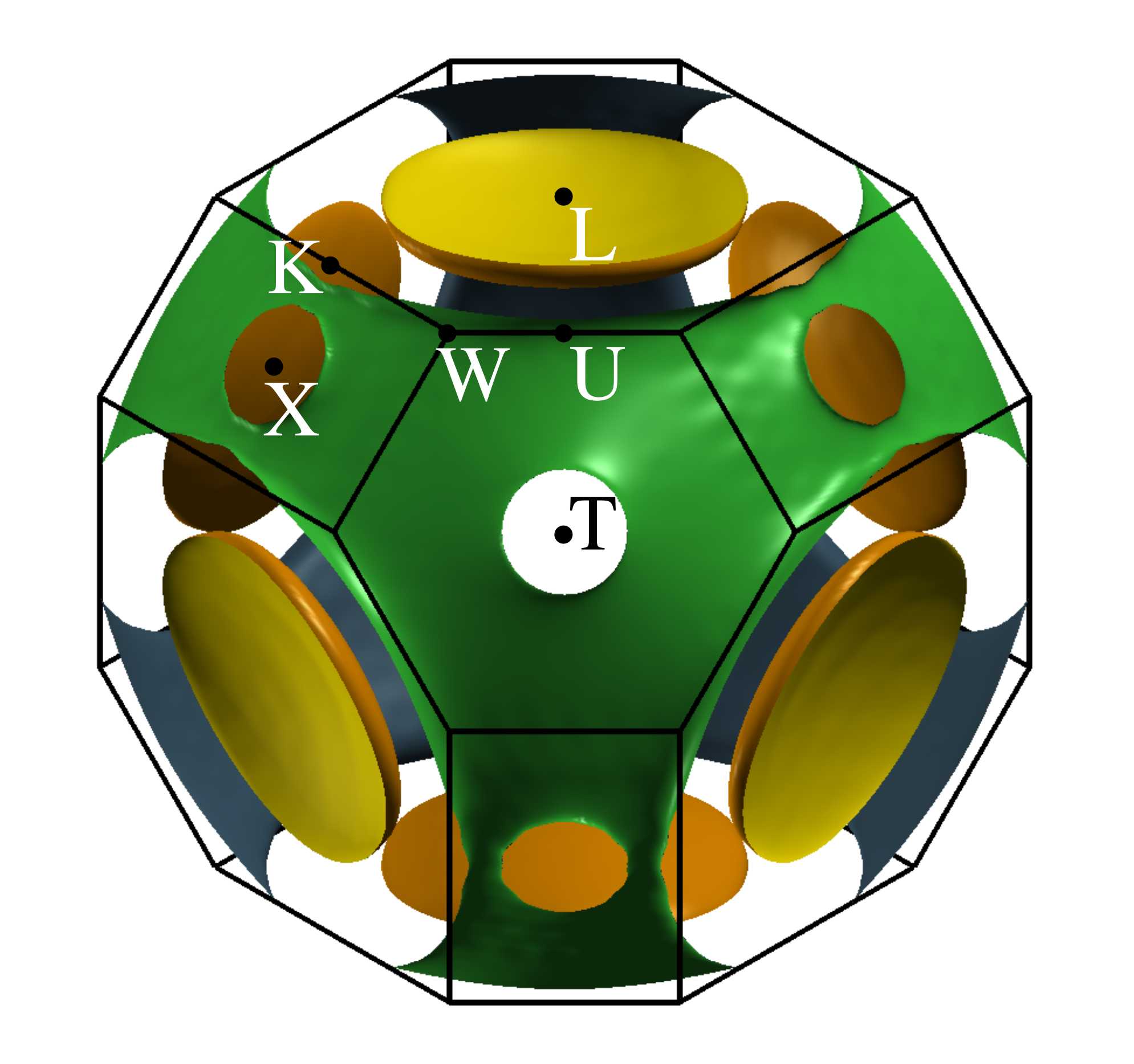}\includegraphics[width=0.575\linewidth]{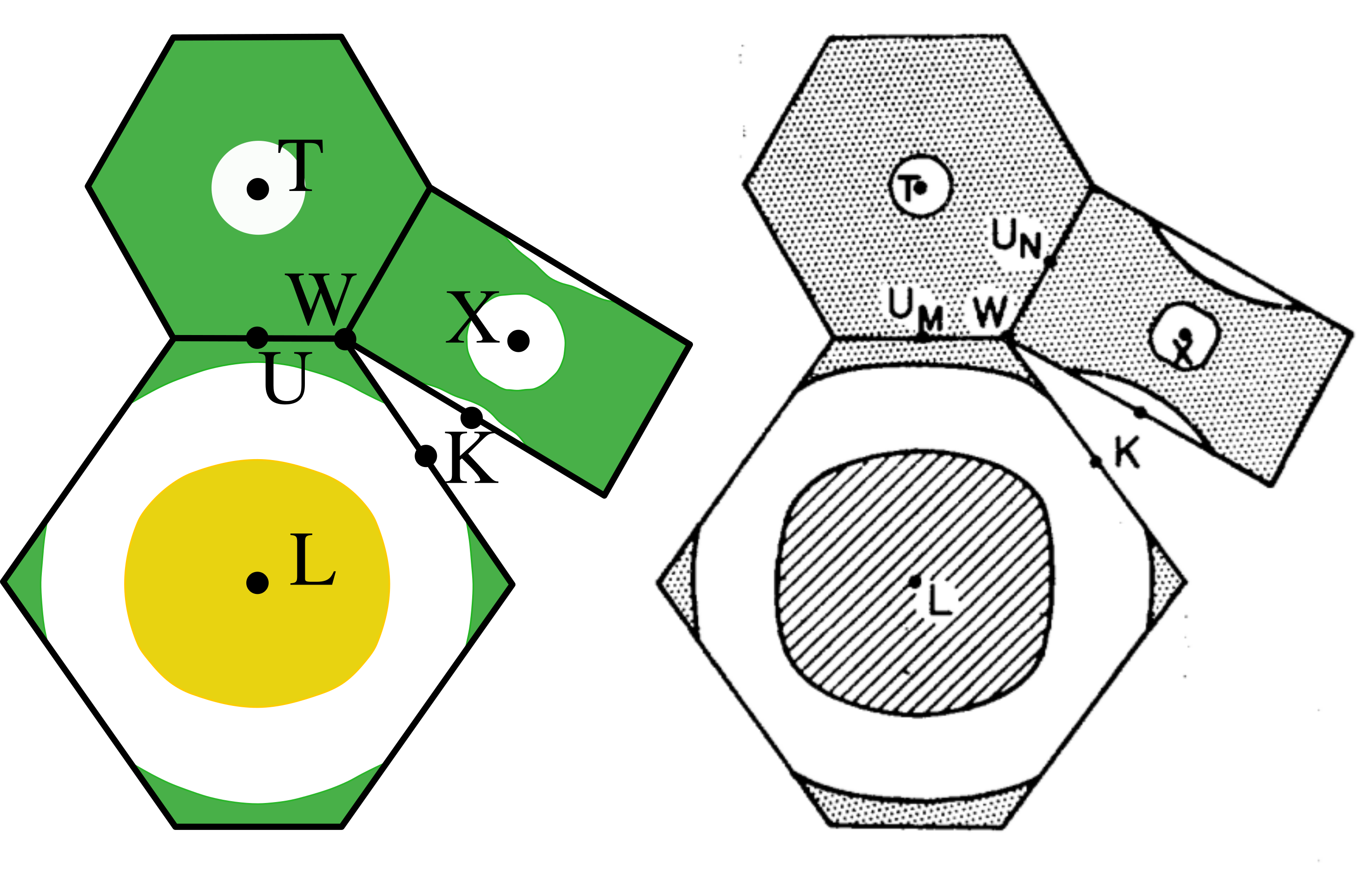}
\caption{(Top panel) Electronic band structure and density of states (DOS) of $\alpha$-Hg, with (continuous line) and without (dashed line) spin-orbit coupling. (Bottom panel) from left to right, 3D plot of the Fermi surface from fully relativistic calculations; 2D cuts along the BZ 
boundary are compared to
the corresponding experimental cuts from Ref.~\onlinecite{PhysRev.152.548}.}\label{fig2}
\end{figure}

{\bf Electronic structure.}
We start  
from the electronic band structure, shown in Fig.\ref{fig2}.
In agreement with previous literature,\cite{PhysRev.152.548,4bddb6a15dc748f89975d64fc087edd7,doi:10.1002}
we find a well-dispersed parabolic band, derived from $s$-states,  partially hybridized with unoccupied $p$ states. 
In the region between 5.5 eV and 9~eV below the Fermi level ($E_F$),
the $s$-parabola is tangled with the $d$-states.   
Including relativistic spin-orbit coupling (SOC) 
causes sizable effects in the $d$ band region 
and, to a lesser extent, in the vicinity of $E_F$.
In particular, SOC removes several band degeneracies,
for example around the $L$ point and along the $K\rightarrow X$ path,
-- compare full and dashed lines in Fig.~\ref{fig2}.

The resulting density of states (DOS) has a rather interesting shape: 
a broad feature, corresponding to $s$-states, extends from $\simeq -10$ eV to $E_F$, and two high, narrow peaks, due to the two groups of spin-orbit splited $d$ bands, are centered around 6 and 9 eV below $E_F$.

To the best of our knowledge, the band structure of mercury has never been measured by Angle-Resolved-Photo-Emission Spectroscopy, but
indirect evidence of the Fermi surface shape can be inferred from 
de-Haas-van-Alphen, magneto-resistance, and cyclotron-resonance measurements\cite{BRANDT196518,PhysRev.152.548,doi:10.1002}.
In the bottom-left panel of Fig.\ref{fig2}, we show a three-dimensional view of the calculated Fermi Surface,
which comprises two disconnected parts: a tubular network extending throughout the Brillouin zone (BZ),
and a disk enclosing the $L$-point. 
Our calculations reporoduce the experimental measurements
with striking accuracy:
not only the main features, but also finer details, such as the small circular hole pockets around the $X$ and $T$ points, and an elongated hole pocket around the $K$-point, are perfectly reproduced.

\begin{figure*}[]
\centering
\includegraphics[width=0.80\linewidth]{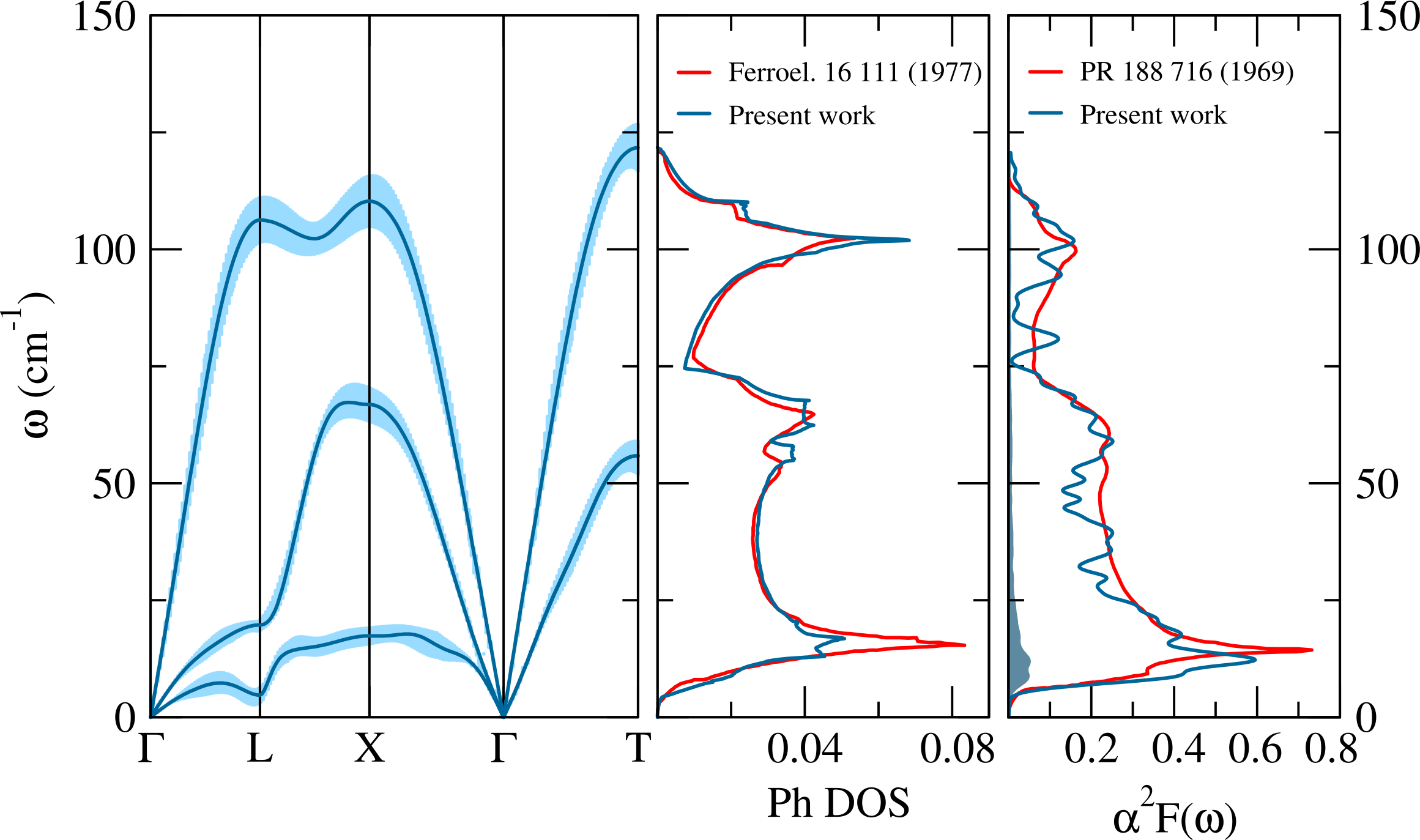}
\caption{Vibrational frequencies of $\alpha$-Hg. We highlight the role of spin-orbit coupling reporting both results with (continuous line) and without (dashed line) the SOC contribution. For comparison the experimental measurements from Ref.\citenum{doi:10.1080/00150197708237132} are reported as empty black circles.
}\label{fig3}
\end{figure*}%
{\bf Vibrational Properties.}
Besides the low-energy electronic structure, DFT-GGA calculations also reproduce
with excellent accuracy the phonon dispersions, provided that spin-orbit coupling is included and the experimental structure is considered.

In Fig.~\ref{fig3}, calculated phonon dispersions 
are compared to the neutron scattering data from Ref.~\onlinecite{doi:10.1080/00150197708237132}.

The phonon spectrum extends up to 120~cm$^{-1}$ with  a pseudo-gap around 75~cm$^{-1}$ separating transverse and longitudinal modes.
The lower transverse branch is very soft and almost flat throughout the whole BZ.
Around the  $L$-point, a further softening occurs: here
$\omega_L$ is only $6.5$~cm$^{-1}$. We find that including relativistic (SOC) effects is crucial for correctly capturing the experimental dispersion in the low-frequency region and obtaining a dynamically stable structure. In fact, without SOC the frequencies of the transverse branch around the $L$ point are imaginary 
(Fig.~\ref{fig3}).
This result is 
consistent
with the recent report that relativistic effects are required to explain also the low melting temperature of mercury~\cite{melting}.

\begin{figure}[]
\centering
\includegraphics[width=0.9\linewidth]{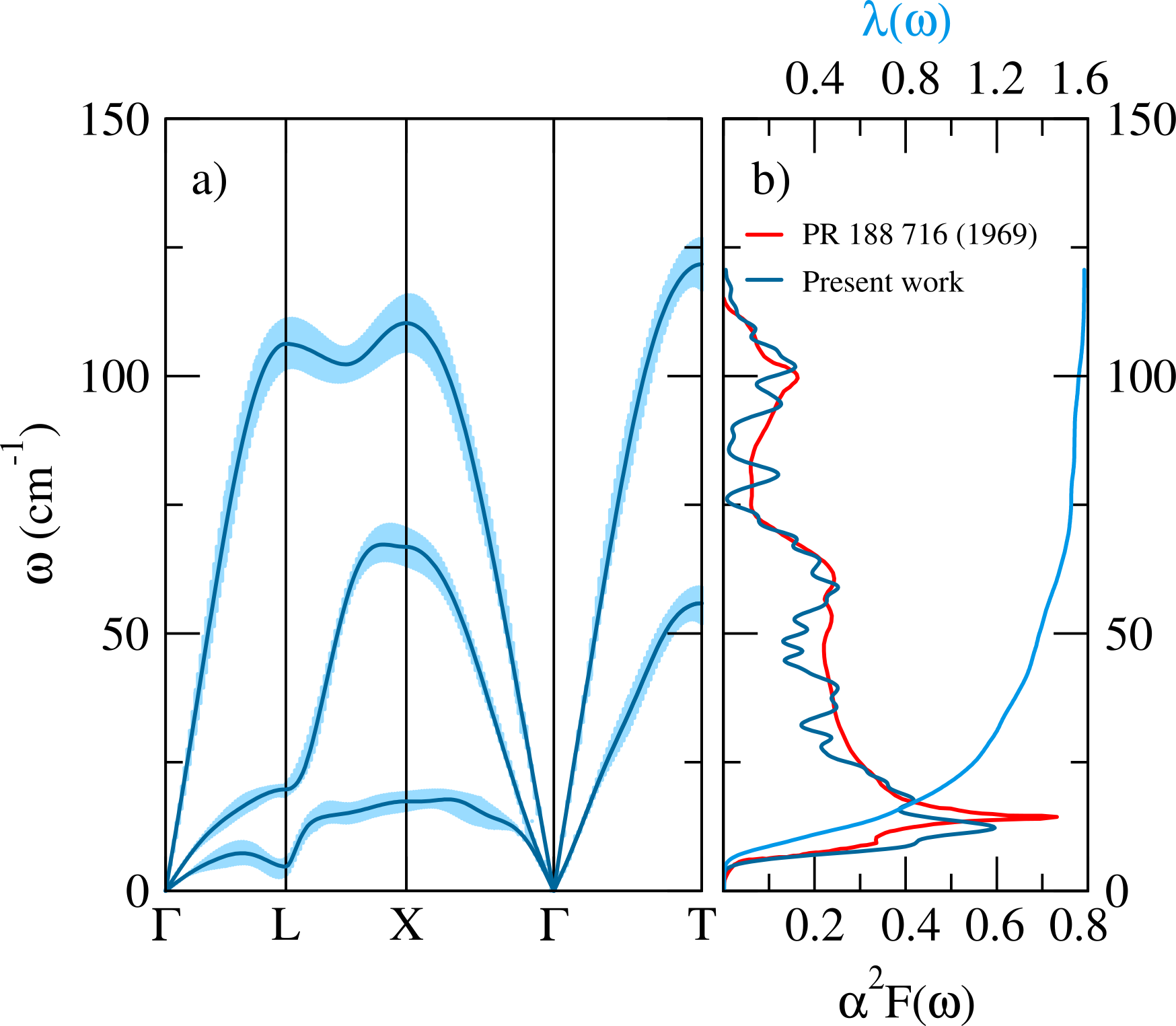}
\caption{Panel (a) Phonon dispersion. The thickness of the light-blue shading is proportional to the phonon linewidths.\\
  Panel (b), Calculated \'Eliashberg function (blue), and $\lambda(\omega)$ (light-blue, top side scale). 
  Experimental $\alpha^2F(\omega)$ from tunneling (Ref.~\onlinecite{PhysRev.188.716}), red. 
}\label{fig4}
\end{figure}

{\bf Electron-phonon coupling.}
In the left panel of Fig.~\ref{fig4} we report the branch- and momentum-dependence of the 
electron-phonon linewidth $\gamma_{\mathbf{q}\nu}$ (half-width at half-maximum):
\begin{equation}
\gamma_{\mathbf{q}\nu}=\frac{2\pi \omega_{\mathbf{q}\nu}}{N_k} \sum_{\mathbf{k}nm}
|g^{\mathbf{q}\nu}_{\mathbf{k+q}m,\mathbf{k}n}|^2 \delta(\varepsilon_{\mathbf{k+q}m}-\varepsilon_F)\delta(\varepsilon_{\mathbf{k}n}-\varepsilon_F)
\label{gamma}
\end{equation}

In Eq.\ref{gamma} the summation of electron-phonon matrix elements, $g^{\nu}_{\mathbf{k+q}n,\mathbf{k}m}$\cite{PhysRevLett.101.016401}, is performed on electronic states ($\varepsilon_{\mathbf{k}n}$) at the Fermi level ($\varepsilon_F$) using $N_k$ $k$-points in the BZ. 
The phonon linewidths, and hence the electron-phonon coupling,
are rather constant over all phonon modes. However, 
the presence of a 
 soft and weakly dispersive phonon branch causes a  pronounced peak at about 15~cm$^{-1}$ in the phonon DOS and in the \'Eliashberg electron-phonon coupling spectral function (Fig.~\ref{fig4}b): 
\begin{equation}
\alpha^2 F(\omega)=\frac{1}{2\pi N(\varepsilon_F)N_q}\sum_{\mathbf{q}\nu}
\frac{\gamma_{\mathbf{q}\nu}}{\hbar \omega_{\mathbf{q}\nu}}\delta(\omega-\omega_{\mathbf{q}\nu}),
\end{equation}
obtained summing over $N_q$ phonons with  wavevectors $\mathbf{q}$ and mode index ($\nu$) with frequency 
$\omega_{\mathbf{q}\nu}$.
The shape of the $\alpha^2 F(\omega)$ results  in a large electron-phonon coupling parameter $\lambda(\omega\to\infty)=2\int_{0}^{\omega}\alpha^2F(\omega')/\omega' d\omega'$ = 1.57 
and a rather small logarithmic-averaged phonon frequency: $\omega_{\textrm{log}}=27.3$~K.

The agreement with tunneling measurements\cite{doi:10.1080/00150197708237132} is excellent: both the soft-phonon peak at low frequencies (below 25~cm$^{-1}$),
and the reduced coupling of the longitudinal mode at higher frequencies are well reproduced by our calculations.
The calculated $\omega_{\textrm{log}}$ and $\lambda$ are also in agreement with the corresponding experimental values from tunneling ($\omega_{\textrm{log}}=29$~K  
 and $\lambda$=1.6~\cite{PhysRev.188.716,PhysRevB.12.905}) and with the  specific-heat renormalization data for $\lambda$~\footnote{We derived the $\lambda$ parameter from the experimental Sommerfield coefficients $\gamma$ according to the following relation: $\lambda^{exp}=\frac{\gamma^{exp}}{\gamma^{th}}(1-\lambda^{th})-1$  } ($\lambda^{sh}\sim$1.56\cite{PhysRev.135.A631},1.66\cite{RevModPhys.36.131},1.58\cite{PhysRev.140.A507} and $\sim 2.0$~\cite{PhysRev.118.127}).

{\bf Superconducting Properties.}
Superconducting Density Functional Theory (SCDFT),~\cite{Davydov_EliashbergVsSCDFT2020} is  an extension of DFT to the SC phase, which was developed  
with the explicit purpose\cite{PhysRevLett.60.2430,PhysRevB.72.024545,PhysRevB.72.024546,PhysRevLett.94.037004} of treating both the electron-phonon and the Coulomb interaction on an equal footing, eliminating any adjustable  parameters, such as the empirical Coulomb pseudopotential $\mu^*$.

\begin{figure}[]
\centering
\includegraphics[width=0.8\linewidth]{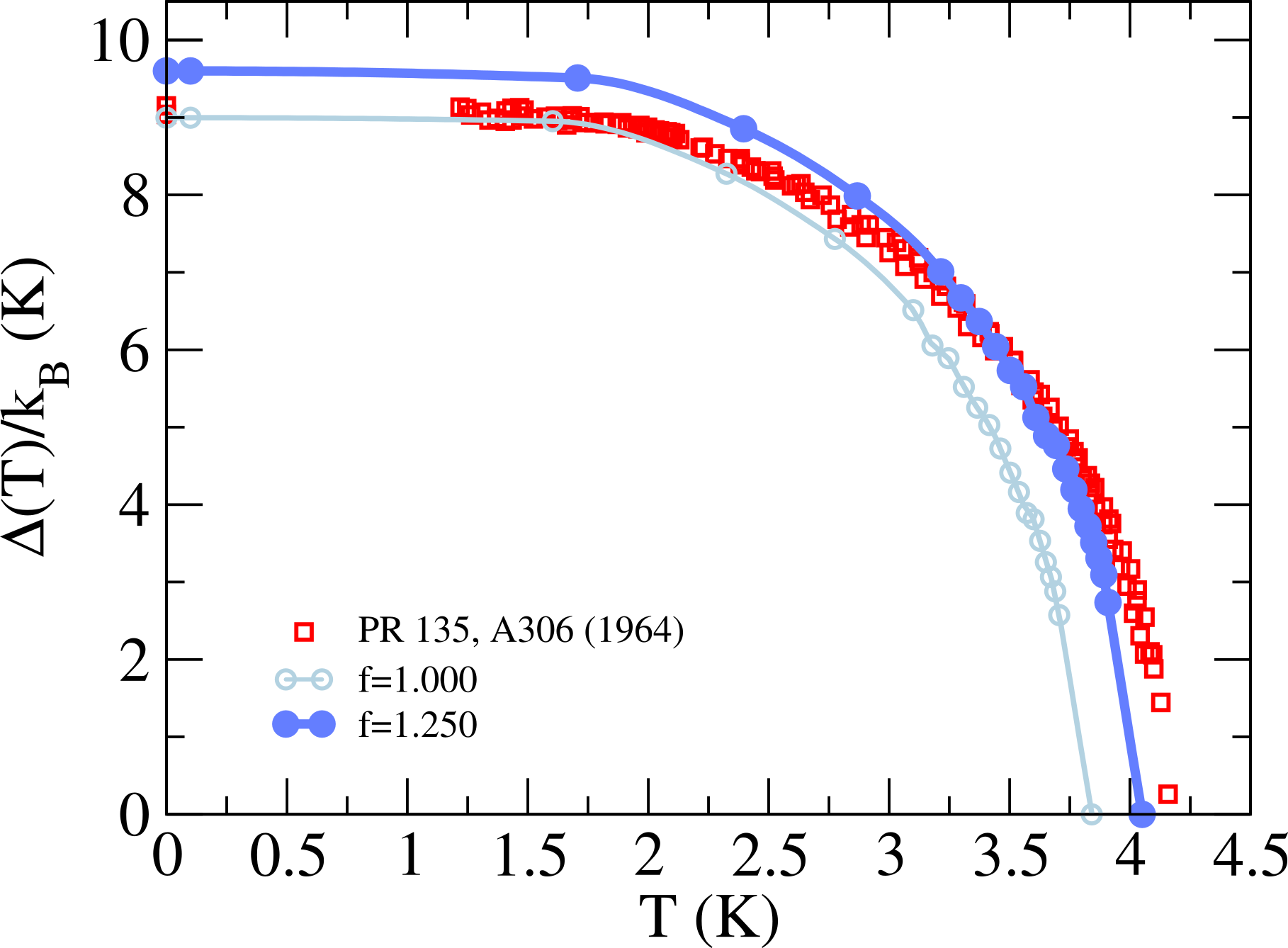}
\caption{Comparison between theoretical (linearly connected light blue circles) and experimental gap\cite{PhysRev.135.A306} (red squares) plotted as a function of temperature. The blue (linearly connected) points  are obtained with a 1.25 energy scaling  as described in the main text.
}\label{fig7}
\end{figure}

The solution of the SCDFT gap equation\cite{SPG_SCDFTfunctional_PRL2020,FloresLivas2020} for $\alpha$-Hg in the static and isotropic approximation, including both electron-phonon and electron-electron interactions, reproduces experimental
data with remarkable accuracy, as shown in Fig.~\ref{fig7}, where the temperature dependence of the SC gap (at $E_F$) obtained in SCDFT (light blue open circles) is
compared with tunneling data from Ref.~\onlinecite{PhysRev.135.A306} (red squares).

The two curves follow each other rather closely.
The critical temperature obtained extrapolating the calculated low-T
data is \Tc$^{\rm SCDFT}$=3.84~K, to be compared with the experimental
value \Tc$^{\rm exp}$=4.15~K. For the BCS ratio $2\Delta(0)/k_B$T$_{\rm C}$,
SCDFT predicts a value of 4.70, to be compared with experimental values  of 4.6$\pm 0.2$\cite{PhysRev.119.575} and 4.60$\pm 0.11$\cite{PhysRev.135.A306}. %
This value places Hg in the strong-coupling regime; the 
 low \Tc\  results essentially from the extremely low phonon frequencies. 

{\bf Coulomb Interaction.}
In addition to the electron-phonon interaction, SCDFT 
gives a microscopic insight also into the residual Coulomb scattering,
an aspect disregarded in most studies of SC, which typically employ
the popular Morel-Anderson approximation.
Also this aspect is strongly anomalous in mercury, as we will show in the following.

In SCDFT, in the absence of SOC, the Coulomb interaction between electrons is described by
the iso-energy surface average $V(\varepsilon, \varepsilon^\prime)$ of the screened Coulomb matrix elements $V_{i\mathbf{k},j\mathbf{k}^\prime}$:\cite{Sanna2018_EliashbergJPSJ,Massidda_SUST_CoulombSCDFT_2009} 
\begin{equation}\label{coul}
V(\varepsilon, \varepsilon^\prime)=\sum_{i\mathbf{k},j\mathbf{k^\prime}}V_{i\mathbf{k},j\mathbf{k}^\prime}
\frac{\delta(\varepsilon-\varepsilon_{i\mathbf{k}})}{N(\varepsilon)}
\frac{\delta(\varepsilon^\prime-\varepsilon_{j\mathbf{k}\prime})}{N(\varepsilon^\prime)}
\end{equation}

The effect of $V(\varepsilon, \varepsilon^\prime)$ 
depends crucially on the energies $\varepsilon, \varepsilon^\prime$ of the two electrons involved in the SC pairing: Coulomb interaction will in fact {\em suppress} superconductivity if both states lie in energy regions where the SC gap is positive\cite{MorelAnderson,PhysRev.148.263}, i.e. close to $E_F$, but can also {\em favor} it, if one of the two electrons occupies a state at high energies, 
where the SC gap is negative. In this case, high-energy states will cause a net renormalization (reduction) of the effective Coulomb interaction ~\cite{FloresLivas2020}.

\begin{figure}[]
\centering
\includegraphics[width=1.05\linewidth]{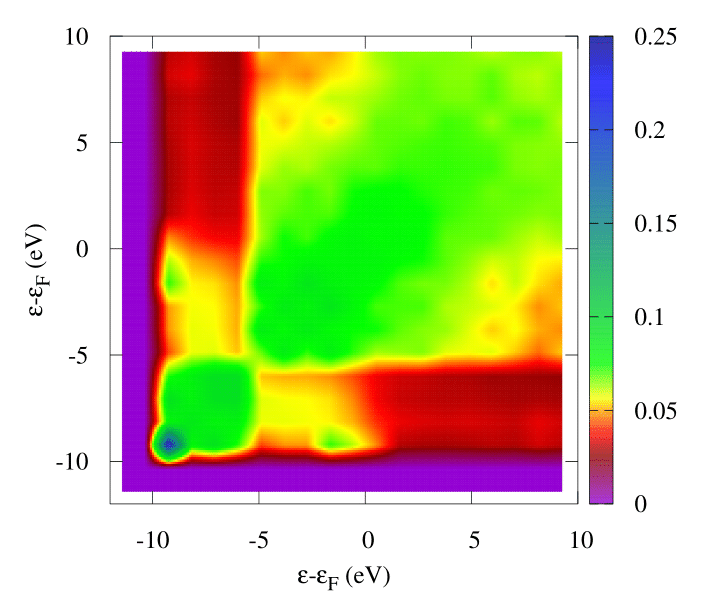}
\caption{Two-dimensional plot of the Coulomb potential as a function of the energy with respect to the Fermi level. The color scale expresses the intensity of $V(\varepsilon,\varepsilon^{\prime})$ in Ry.\\
}\label{fig5}
\end{figure}

A two-dimensional plot of the calculated $V(\varepsilon, \varepsilon^\prime)$ function for mercury is
shown in Fig.\ref{fig5} -- here and in the following, energies are measured from $E_F$.
 Due to the different nature and dispersion of the $s$ and $d$ bands the diagonal elements of $V(\varepsilon, \varepsilon^{\prime})$ show an hot-spot (blue) around $-10$~eV,
corresponding to the bottom of the $s$ parabola in Fig.~\ref{fig1}, 
a square-like feature, with moderate coupling (green) from -8 to -5 eV, 
related to $d$ states,   and an extended region from zero to high energies (green), related to  $s$ and $p$ states. The off-diagonal $\varepsilon=0$ row,  $V(0, \varepsilon^\prime)$, which accounts for interband contributions involving the $s$ states at the Fermi level and all the other states, is non-zero in the low-energy $s$ region,
is very low (nearly zero) for $d$ states ($\varepsilon^\prime < -5$~eV), and  different from zero again only for $\varepsilon^\prime \sim -10$ eV, i.e. near the bottom of the $s$ parabola in Fig.~\ref{fig1}.

Based on this energy structure, we expect that the net effect of  Coulomb interactions on \Tc\; will be rather weak, due both to low diagonal matrix elements in the low-energy (repulsive) region, and large  inter- and intra-band contributions in the high-energy (attractive) regions.

The {\em diagonal} part of the $V(\varepsilon, \varepsilon^{\prime})$ kernel, evaluated at the Fermi energy ($\varepsilon=0$), yields the so-called $\mu$ parameter. For mercury the calculated  $\mu=0.159$ is in line with its neighbours in the periodic table, like Au and Cd ($\mu=$0.136 and 0.142, respectively\cite{PhysRevB.101.134511}), but much smaller than 
the average value $\mu=0.25$ found in most other elemental superconductors like Pb or Al ~\cite{PhysRevB.101.134511}. 
Together with the large bandwidth of the $s$ band, and the extremely small characteristic frequency of mercury, this translates into a Morel-Anderson pseudopotential $\mu^*$=0.07, significantly smaller than the standard value $\mu^*=0.10$ -- details in the Supplementary Material.

However, the most interesting anomaly in the Coulomb screening, which cannot be captured by the standard Morel-Anderson approach, is
connected to the {\em off-diagonal} part of the $V(\varepsilon, \varepsilon^{\prime})$ kernel.

{\bf Influence of $d$ states on \Tc.} 
Due to the presence of non-negligible off-diagonal $s-d$ Coulomb matrix elements, the calculated \Tc\; in mercury turns out to depend
 in a critical way on the position of the high-energy $d$-states.
 
\begin{figure}[]
\centering
\includegraphics[width=0.95\linewidth]{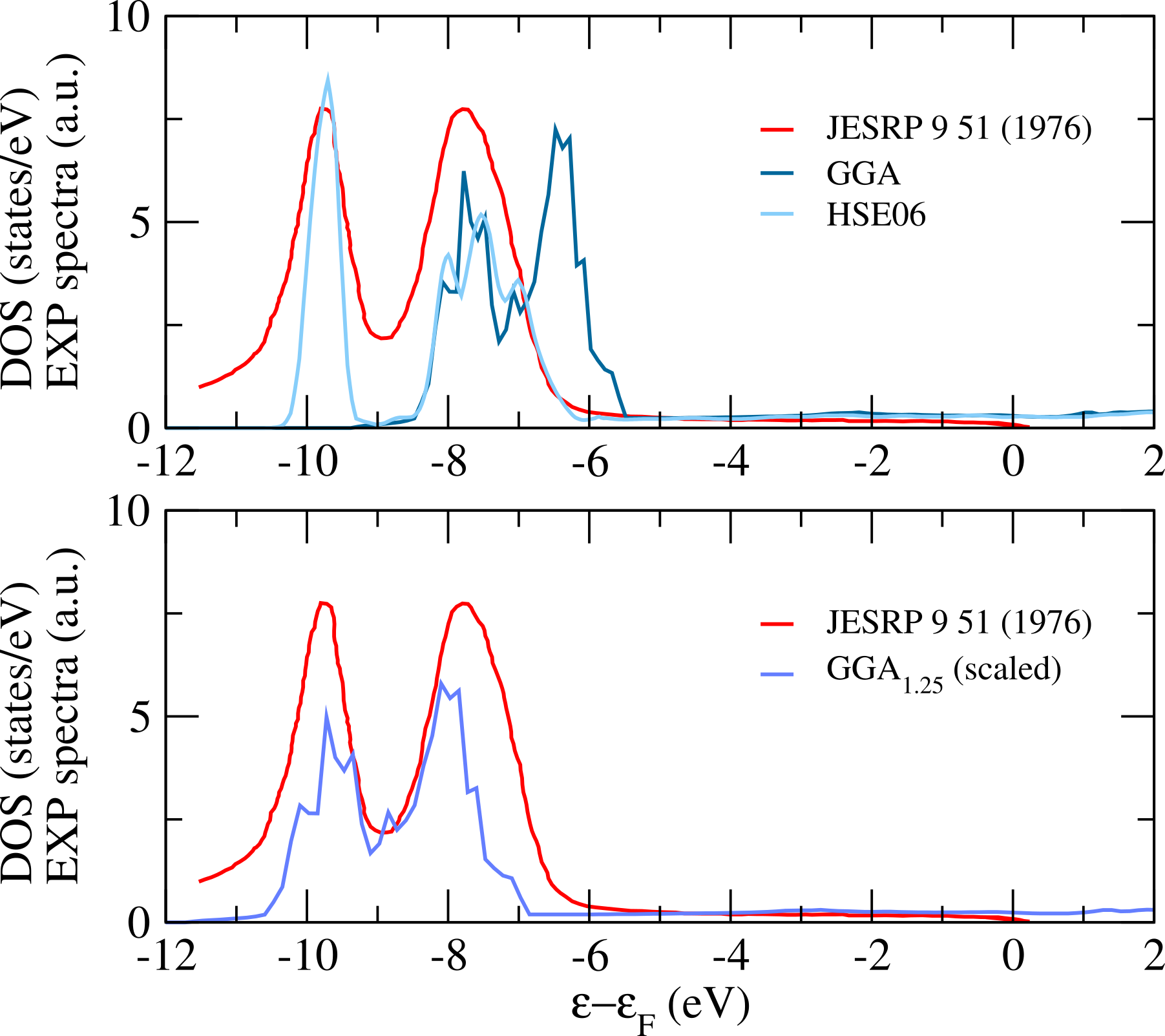}
\caption{Top panel: Electronic spectra calculated in GGA (blue) and hybrid HSE06\cite{HSE1, HSE2} (light blue), and X-Ray Experiments from
  Ref.~\onlinecite{SVENSSON197651}.(red) Bottom panel: Linearly-scaled GGA DOS (f$=1.25$).
\label{fig_6}}
\end{figure}

 This was verified through a simple {\em gedanken experiment}, in which
 we solved again the SCDFT equations, leaving all terms unchanged, apart
 from a scaling of the electronic spectrum, necessary
 to bring the energy position of the calculated DFT-GGA-SOC $d$ bands 
 with experimental X-Ray photoemission data.\cite{SVENSSON197651}
 The physical origin of the energy shift between DFT-GGA calculations
 and experiments is the lack of non-local exchange and correlation terms \cite{PhysRevB.86.125125,PhysRevB.66.115101,PhysRevB.84.205205}; in fact,
 the shift can be easily removed employing non-local functionals, such as the  Heyd-Scuseria-Ernzerhof (HSE06) functional \cite{HSE1, HSE2} -- upper panel of Fig.~\ref{fig_6}.

  As shown in the lower panel of the same figure, a simple linear scaling
  of the whole DFT-GGA-SOC spectrum f$=1.25$ 
is sufficient to mimick this effect and produce an almost perfect agreement between spectra and with experiments.

Solving the SCDFT equations as a function of temperature, we obtain 
the data shown as blue filled circles in Fig.\ref{fig7}: the 
\Tc, obtained from extrapolation of the low-T data, is 4.05~K, with a clear improvement with respect to our previous  GGA-SOC result (see Supplementary Materials). Although the almost perfect agreement with experiment may be fortuitous,
this numerical experiment demonstrates that
shifting the position of apparently inert high-energy states can lead to
a 10$\%$ effect on \Tc.

\section{Conclusions}

In this work we carried out a critical study 
of the superconducting properties of $\alpha$-mercury, aimed at understanding whether this material, which played an essential role in superconductivity history,
can be described by state-of-the-art computational methods. 
Our first-principles calculations, validated with an extensive comparison with available experimental literature, demonstrate that state-of-the-art 
 SCDFT can 
 describe the superconducting state of Hg,
 provided that special care is taken to handle several anomalous electronic
 and lattice properties.

In particular, ($i$) due to strong 
 non-local exchange and correlation effects structural properties are so poorly described by standard density functional, that in order to obtain meaningful electronic and phonon spectra, all calculations have to be performed in the experimenetal crystal structure; ($ii$) SOC effects are also extremely strong, and crucially affect dynamical stability; ($iii$) due to anomalously large off-diagonal $s-d$ matrix elements,  the effective Coulomb potential is
 strongly affected by the energy position of the low-energy $d$ states.
 Taken as a whole, our results demonstrate that, even for an apparently simple compound like mercury, common approximations cannot be applied blindly,
 as this may cause severe qualitative and quantitative errors. This
 aspect is crucial for future high-throughput calculations.
 We would also like to stress that some of the effects discussed here, may appear spectacularly enhanced in high-\Tc\; conventional superconductors, such as the recently discovered superhydrides,
where renormalization of the Coulomb interaction has been invoked to justify differences as large as 100 K in the calculated \Tc's~\cite{https://doi.org/10.1002/adma.202006832}.

\section*{Appendix: Computational details}
All calculations were performed using the plane-wave pseudopotential DFT \textsc{Quantum-Espresso} package~\cite{QEcode, QE-2017} including relativistic effects. 
We used  Optimized Norm-Conserving Vanderbilt Pseudopotentials\cite{PhysRevB.88.085117,PhysRevB.95.239906,VANSETTEN201839} including 5$s$, 5$p$, 5$d$ and 6$s$ states in valence, and the Generalized Gradient Approximation (GGA) for the Exchange and correlation term, with an energy cut-off of 70~Ry.

Integrations over the BZ 
were carried out using uniform 18$\times$18$\times$18 Monkhorst and Pack grids\cite{PhysRevB.13.5188} and
a 0.02~Ry Gaussian smearing.

Phonon frequencies and electron-phonon matrix elements were calculated using linear response theory~\cite{QEcode, QE-2017}, on a 8$\times$8$\times$8 grid to which correspond 65 $q$-points in the irreducible BZ and a dense 24$\times$24$\times$24 mesh for electronic wavevectors.

Total electron-phonon coupling parameter is calculated Wannier interpolating the electron-phonon matrix elements\cite{MarzariComposite,MarzariEntangled,PhysRevB.82.165111} on a denser phononic and electronic meshes of $12\times12\times12$ and $36\times36\times36$, respectively.

The SC critical temperature mediated by electron-phonon interaction has then been calculated fully {\it ab-initio} in the SCDFT framework\cite{PhysRevLett.60.2430,PhysRevB.72.024545,PhysRevB.72.024546,PhysRevLett.94.037004}, using the most accurate available functional~\cite{SPG_SCDFTfunctional_PRL2020}. In this picture, the Coulomb interaction is treated self-consistently at the same level 
as the electron-phonon interaction, in the static isotropic approximation. 
Screened Coulomb matrix elements were calculated in the random phase approximation as in Ref.\onlinecite{Sanna2018_EliashbergJPSJ,PhysRevMaterials.3.114803} without relativistic effects (being the spin-orbit effects negligible around the Fermi energy).

The HSE06 DOS in Fig.\ref{fig7} was obtained using the VASP\cite{Kresse199615,PhysRevB.54.11169,PhysRevB.59.1758} code.

\section*{Data availability}
All the data that support the findings of this study are available from the corresponding authors (C.T. and G.P.) upon reasonable request.

\newpage

\bibliography{bibliography}{}

\section*{Acknowledgements}
L. B., C. T. and G. B. Bachelet acknowledge support from Bando Ateneo Sapienza, 2017-2020.
G. P. acknowledges financial support from the Italian Ministry for Research and Education through PRIN-2017 project ``Tuning and understanding Quantum phases in 2D materials - Quantum 2D" (IT-MIUR Grant No. 2017Z8TS5B).

\end{document}